\documentstyle[prb,twocolumn,aps]{revtex}
\tighten
\begin{document}

\title{Chaotic dynamics of electric-field domains in periodically driven
superlattices}
\author{O. M. Bulashenko\cite{byline}, M. J. Garc\'{\i}a, and L. L. Bonilla}
\address{Universidad Carlos III de Madrid, Escuela Polit\'{e}cnica Superior,
Butarque 15, 28911 Legan\'{e}s, Spain }
\address{\rm (11 October 1995)}
\address{\mbox{ }}
\address{\parbox{14cm}{\rm \mbox{ }\mbox{ }\mbox{ }
Self-sustained time-dependent current oscillations under dc voltage bias have
been observed in recent experiments on $n$-doped semiconductor superlattices
with sequential resonant tunneling.
The current oscillations are caused by the motion and recycling of the domain
wall separating low- and high-electric-field regions of the superlattice, as
the analysis of a discrete drift model shows and experimental evidence
supports.
Numerical simulation shows that different nonlinear dynamical regimes of the
domain wall appear when an external microwave signal is superimposed on the dc
bias and its driving frequency and driving amplitude vary.
On the frequency -- amplitude parameter plane, there are regions of entrainment
and quasiperiodicity forming Arnol'd tongues. Chaos is demonstrated to appear
at
the boundaries of the tongues and in the regions where they overlap.
Coexistence of up to four electric-field domains randomly nucleated in space is
detected under ac$+$dc driving.
}}
\address{\mbox{ }}
\address{\parbox{14cm}{\rm PACS numbers:  73.20.Dx, 73.40.Gk, 47.52.+j}}

\maketitle

\makeatletter
\global\@specialpagefalse
\def\@oddhead{\underline{Submitted to Physical Review B\hspace{281pt}
arch-ive/9510126}}
\let\@evenhead\@oddhead
\makeatother

\narrowtext

\section{Introduction}
Negative differential conductivity (NDC) in weakly-coupled narrow-miniband
semiconductor superlattices (SL's) results in the formation of electric-field
domains, which have been studied both experimentally \cite{domains-exp} and
theoretically.  \cite{millaikh,KAL,Prengel,Bonilla94,NATO}
The domains are stable if the doping or photoexcitation are large enough to
form
a stationary charge accumulation layer (the domain wall). The domain wall moves
from well to well as the bias increases and gives rise to the jumps
(discontinuities) of the current in the stationary $I$-$V$ characteristics.

When the carrier density is not sufficiently large to form stable domains, but
it is large enough for the uniform field distribution to be unstable, periodic
time-dependent oscillations of the current under fixed dc bias are possible.
These oscillations have been observed in recent experiments on GaAs/AlAs SL's;
they are damped for undoped photoexcited samples \cite{Kwok} and undamped for
doped SL's without photoexcitation. \cite{japan,Kastrup}
According to a discrete drift model, \cite{Bonilla94} the current oscillations
are caused by a periodic motion of the domain wall over a few periods of the
SL,
\cite{Bonilla-ICPF}, and this is confirmed by photoluminescence measurements.
\cite{japan,Kastrup} Note that damped oscillations (with frequencies up to 20
GHz) have been observed experimentally by Le Person {\it et al.}
\cite{leperson}
in a photoexcited wide-miniband SL with strong interwell coupling.
These oscillations have been interpreted in terms of dipole charge waves.
We do not consider the miniband transport regime in the present paper.

This situation is reminiscent of that found in bulk GaAs with NDC caused by
the intervalley transfer of electrons, where the well-known Gunn oscillations
mediated by high-field domain dynamics may appear under dc voltage bias.
\cite{shaw,higuera92} There are several important differences between the Gunn
oscillations and those observed in the SL's.
A significant difference is that the space charge wave is a $dipole$ in the
case
of the Gunn oscillations (a domain of high electric field with $two$ charge
layers: accumulation and depletion) and a charge $monopole$ in the SL current
oscillations (a wave of $one$ charge accumulation layer, or the domain wall,
connecting two electric-field domains).
Another difference is that the Gunn waves are generated close to the injecting
contact whereas the domain walls appear clearly inside the SL
\cite{Bonilla-ICPF}.  Notice finally that the different transport mechanisms
determining the velocity of waves and the characteristic frequency of the
oscillations give rise to different limitations in the performance of possible
devices.
While for the Gunn effect the oscillation frequency is limited by a parameter
of the material (the intervalley scattering time), for the SL's it is
determined
by the tunneling time. Hence, the oscillation frequency can be varied by tuning
the growing parameters (barrier widths, etc.) and/or the dc bias.
The bias regions between different resonance peaks give rise to quite different
frequencies, ranging between hundreds of KHz and several GHz over wide
temperature ranges including room temperature. \cite{Kastrup}

In our previous paper \cite{BB95} we predicted the appearance of chaos by
applying an additional small ac signal to the dc voltage bias in the regime
where the SL exhibits self-sustained oscillations.
In particular, when the ratio between the natural and the driving frequencies
is
fixed at the golden mean, chaotic current oscillations and strange attractors
with a multifractal measure have been obtained. \cite{BB95}

Here we present a detailed numerical study of nonlinear current oscillations in
SL's over the whole driving frequency -- driving amplitude parameter plane.
Based upon a relatively simple (but otherwise self-consistent) discrete drift
model, \cite{Bonilla94} the shape of frequency-locked regions (Arnol'd
tongues),
chaotic and quasiperiodic regions is determined, thereby providing a global
bifurcation picture. The largest Lyapunov exponents are calculated to identify
chaotic solutions and to characterize their fractal dimension.
The critical line where the Arnol'd tongues begin to overlap and chaos appears
is found at very low values of the amplitude of the driving force, unlike other
periodically driven semiconductor systems which also
display complex nonlinear interaction between internally generated oscillations
and an external ac signal (e.g., GaAs Gunn diodes, \cite{mosekilde} $p$-Ge,
\cite{kahn92,BT}).

An additional point worth noticing is that the physical mechanism of the NDC in
the SL system (responsible for chaos) is based on a pure quantum-mechanical
effect (resonant tunneling) which is absent in the classical limit.
Thus we may have a case of quantum chaos without classical counterpart
\cite{capasso92} which might be easier to detect experimentally than that
proposed by Jona-Lasinio {\it et al.} \cite{capasso92}. In both cases the
self-consistent mean-field potential created by the charges is instrumental in
inducing chaos, which is also the case when a quantum system is coupled to
purely classical subsystems having a widely different time scale. \cite{BG}
All these phenomena are different from the so-called quantum chaos,
\cite{haake}
i.e., the behavior of quantum systems whose classical counterpart is chaotic.

Recently chaotic dynamics in semiconductor superlattices with miniband
transport
has also been considered within a classical balance-equation approach.
\cite{aleks} In this case the chaotic dynamics is due to the negative
effective mass of the electrons in the minibands for the regime of the Bloch
oscillations under an external ac electric field (field self-consistency is
also
taken into account). In this completely different physical situation, the field
was assumed to be spatially homogeneous \cite{aleks} which precludes
consideration of the spatial structure of chaos, in contrast with our work.
\cite{BB95}

The paper is organized as follows. In Sec.\ II we describe the physical model
and its governing dynamical equations for a dark doped SL.
In Sec.\ III we recall the main qualitative features of the self-sustained
oscillations in a dc-voltage biased SL.
The general behavior of the self-oscillating SL subjected to external ac and dc
biases is discussed in Sec.\ IV. Frequency-locked solutions (Arnol'd tongues)
are analyzed and compared with those for the Gunn diode.
The chaotic dynamics at the golden mean ratio between the natural and the
driving frequencies is described in Sec.\ V. Different techniques to analyze
chaotic behavior (Poincar\'e mapping, bifurcation diagram, Lyapunov exponents,
phase plots, Fourier spectra, first return map, etc.) are discussed.
Section VI draws the main conclusions of this work.
In the Appendix we present the algorithm to calculate the Lyapunov exponents
for
our system and discuss the Lyapunov dimension calculated by the Kaplan-Yorke
formula.

\section{The physical model}
We consider a set of weakly interacting quantum wells (QW's) under high voltage
biases, so that the electron states are localized in the wells and the
tunneling
process is sequential.
The relaxation from excited levels to the ground state within each QW ($\sim
1$~ps) is considered to be much faster than the tunneling process between
adjacent QW's ($\sim 1$~ns). Therefore, a single QW reaches a local equilibrium
between two consecutive tunneling processes, and the state of the system can be
characterized by values of the electric field ${\cal E}_i(t)$, and the electron
density $n_i(t)$, with $i=1,\dots,N$ denoting the QW index. \cite{Bonilla94}

The one-dimensional equations governing the dynamics of the doped SL are the
Poisson equation averaged over one SL period $l$, Amp\`ere's equation for the
balance of current density, and the voltage bias condition
\cite{BB95}
\begin{eqnarray} \label{pois}
\frac{1}{l} \; ({\cal E}_i - {\cal E}_{i-1}) = \frac{e}{\epsilon} \;
(n_i - N_D), \\   \label{amp}
\epsilon \; \frac{d {\cal E}_i}{d t} + e n_{i} v({\cal E}_i) = J, \\
\label{bias} l \sum_{i=1}^{N} {\cal E}_i = V(t).
\end{eqnarray}
Here $\epsilon$, $e$, and $N_D$ are the average permittivity, the electron
charge, and the average doping density, respectively.
The total current density $J(t)$ is the sum of the displacement current and the
electron flux due to sequential resonant tunneling $e n_i v({\cal E}_i)$.
The effective electron velocity $v({\cal E})$ (proportional to the tunneling
probability) exhibits maxima at the resonant fields for which the adjacent
levels of neighboring QW's are aligned \cite{remark2} (Fig.\ \ref{v-E}).
Notice that the mechanism of sequential resonant tunneling is responsible for
the NDC region in the velocity curve which in turn gives rise to the
current instabilities in the SL.  The voltage $V(t)$ in (\ref{bias}) is the sum
of a dc voltage $V_b$ and an ac microwave signal of relative amplitude $A$ and
driving frequency $f_d$:  $V(t) = V_b \{1+A \sin (2 \pi f_d t) \}$.


The boundary condition at the first contact
$\epsilon ({\cal E}_1 - {\cal E}_0)/(e l) = n_1 - N_D = \delta$ allows for a
small ($\delta \ll N_{D}$) negative charge accumulation in the first well.
The physical origin of $\delta$ is that the $n$-doped SL is typically
sandwiched
between two $n$-doped layers with an excess of electrons, thereby forming a
$n^{+}$-$n$-$n^{+}$ diode. \cite{japan}
Then some charge will be transferred from the contact to the first QW creating
a
small dipole field that will cancel the electron flow caused by the different
concentration of electrons at each side of the first barrier.

Note that the rate equation for the electron density for our model can be
derived by differentiating (\ref{pois}) and using (\ref{amp}):
\begin{equation} \label{n-conserv}
\frac{\partial n_{i}}{\partial t} + \frac{1}{l} \;
[n_{i} v({\cal E}_{i}) - n_{i-1} v({\cal E}_{i-1})] = 0.
\end{equation}
This is the equation of charge conservation under sequential resonant tunneling
between neighboring QW's.

Finally, the current density $J(t)$ in the external circuit under voltage bias
condition can be obtained from the time-derivative of (\ref{bias}) and
Amp\`{e}re's law (\ref{amp}) in the form

\begin{equation} \label{curr}
J(t) = c_V \frac{d V}{d t} + \frac{e}{N} \sum_{j=1}^{N} n_j v({\cal E}_j),
\end{equation}
where $c_V = \epsilon/(l N)$ is the intrinsic SL capacitance per unit
cross-sectional area.

To study our equations, it is convenient to render them dimensionless, using
characteristic physical quantities.
As the unit of the electric field $E={\cal E}/{\cal E}_{1-2}$, we adopt
${\cal E}_{1-2} = \Delta /(e l)$, with $\Delta$ being the energy separation
between the first and the second electron subbands
[the maximum of $v({\cal E})$].
This yields the characteristic charge density
$n_0 = \epsilon \; {\cal E}_{1-2}/(e l)$ used to normalize the doping
$\nu=N_D/n_0$ and the electron density.
The dimensionless velocity ${\rm v}(E)$ is obtained normalizing $v({\cal E})$
by its value at ${\cal E}={\cal E}_{1-2}$, where it has a local maximum due to
resonance in tunneling (see Fig.\ \ref{v-E}).
The others dimensionless quantities are defined as follows:
the time $\tau = t/t_{\rm tun}$ where $t_{\rm tun}=l/v({\cal E}_{1-2})$ is the
characteristic tunneling time; the dc bias ${\cal V} = V_b / {\cal E}_{1-2} l
N$; the ac bias amplitude $a=A {\cal V}$;
the driving frequency $\omega = 2 \pi f_d t_{\rm tun}$.

Now substituting (\ref{pois}) and (\ref{curr}) into (\ref{amp}) we obtain a
system of $N$ equations for the electric-field profiles
\begin{eqnarray} \label{map}
\frac{d E_i}{d \tau} = && \frac{1}{N}\sum_{j=1}^{N} {\rm v}(E_j) \;
[ E_j -E_{j-1} + \nu ] \nonumber \\ &&{}
- {\rm v}(E_i) \; [ E_i -E_{i-1} + \nu ] + a \omega \cos (\omega \tau),
\end{eqnarray}
with the boundary condition $E_0 = E_1 - \nu\delta$ and initial conditions
$E_i(0) = {\cal V},\, \forall i$.

As an example, we consider a GaAs/AlAs SL at $T$=5 K with $N$=40, $l$=13 nm,
$N_D \approx$1.15$\times 10^{17} {\rm cm}^{-3}$,
$\Delta \approx 135$ meV, for which undamped time-dependent oscillations
of the current were first observed. \cite{japan} For these values one gets
${\cal E}_{1-2}\approx$10$^5$ V/cm, $\nu\approx 0.1$, and $t_{\rm tun}\approx$
2.7 ns, and we take ${\cal V}$=1.2 (corresponding to $V_b\approx$7.8 V).

The system (\ref{map}) is solved numerically by the fourth-order Runge-Kutta
method using 4000 time steps per one oscillation period.
We start with a uniform initial field profile and solve the equations for dc
bias. After a short transient, the self-sustained oscillations set in and we
switch on the ac part of the bias.
Every time step we calculate the electric-field distribution over SL $E_i(t)$
and the total current density $J(t)$ [from Eq.(\ref{curr})].
The main features of our numerical results are as follows.

\section{Self-sustained oscillations under dc bias}
\label{selfosc}

For pure dc case ($a$=0) the SL exhibits undamped time-periodic current
oscillations when the doping density is between some critical values $\nu^{*}$
and $\nu^{**}$.
In Fig.\ \ref{j-dop} the temporal behavior of the current starting from $t$=0
is
shown for different doping densities.
Below $\nu^{*} \approx 0.066$ the electric-field distribution over the SL
remains almost uniform, so that the current exhibits no oscillations.
Above $\nu^{**} \approx 0.175$, after some transient period of the order of
$t_{\rm tun} N$, stable stationary electric-field domains are formed. This
results in an increase of the current and its saturation with time.


For in-between values $\nu^{*} < \nu < \nu^{**}$ and appropriate dc voltage
bias, we find undamped oscillations with frequency slightly dependent on the
doping. In Fig.\ \ref{f0-dop} the frequency of the natural oscillations
versus doping concentration is shown for different boundary parameters
$\delta$.
One can estimate the natural frequency to be approximately equal to the inverse
total tunneling time $1/(t_{\rm tun} N)$, which gives, after substitution of
our value for $t_{\rm tun} \approx 2.7$ ns and $N=40$, a frequency about 10 MHz
in close agreement with the value observed in experiment.\cite{japan} This
estimation improves as $N$ increases.
All the results presented below were obtained for the value of
$\delta=10^{-3}$.

The spatial structure of the oscillations can be seen from the electron density
and the electric-field distributions of Fig.\ \ref{En-it}.
Starting from uniform distribution at time $\tau$=0, when the dc bias is
turned on, after a transient period $\tau \approx 3$, a charge accumulation
wave (monopole) is created and then it moves toward the correspondent contact.
Depending on the applied voltage, it may or may not reach the end of the SL
before it disappears and a new monopole is formed starting a new period of the
oscillation.
Simulations clearly show monopole recycling with two monopoles coexisting
during
some part of one current oscillation period [see Fig.\ \ref{En-it}(b)].
The period of the oscillation is mainly determined by the total travelling time
of the monopole across the SL $t_{\rm tun} N$, as was shown in Fig.\
\ref{f0-dop}.


The total current density $J(t)$ determines the average state of the system
(see Eq.\ \ref{curr}), nevertheless the information on the spatial dynamics is
presented in the time dependences of the current.
In the curves of Fig.\ \ref{j-dop} small current spikes can be seen during the
initial stage of the domain formation, which are correspondent to the
well-to-well jumping of the charge accumulation layer.
The similar, but less clear, fine structure can be resolved in the periodic
part
of the current oscillations (become more pronounced after taking the derivative
$dJ/dt$). {}From the number of spikes per period ($\sim$6 for our curves) it is
possible to estimate the number of QW's the monopole moves across.

The existence of the threshold value for the doping $\nu^{*}$, above which
current instabilities take place, can be understood from the charge
conservation
law (\ref{n-conserv}) rewritten in the continuum limit as

\begin{equation} \label{n-cons1}
\frac{\partial n}{\partial t} + \frac{\partial}{\partial x} [n v(E)]=0.
\end{equation}
Taking spatial derivative and using the Poisson law for $\frac{\partial E}{
\partial x}$ one gets
\begin{equation} \label{n-cons2}
\frac{\partial n}{\partial t} + v \frac{\partial n}{\partial x} +
\frac{e}{\epsilon} n v'_E (n-N_D) = 0,
\end{equation}
where the sign of the velocity derivative over the field $v'_E$ is crucial.
In the NDC region where $v'_E < 0$ the last term in (\ref{n-cons2}) gives rise
to an exponential growth of the charge with the characteristic time
$\tau_{\epsilon} \sim \epsilon / (e N_D |v'_E|)$. For $v'_E > 0$ the same term
gives charge relaxation to quasineutrality $n \approx N_D$.
Taking into account the characteristic transit time $\tau_t \sim l N/v$ from
the
convection term of (\ref{n-cons2}) we can get significant growth of charge if
$\tau_t > \tau_{\epsilon}$, that is $N_D N > \epsilon v/(e l |v'_E|)$.
This condition is similar to the critical $nL$-product for the Gunn effect.
\cite{shaw}
The treatment above is of course too rough, but it clarifies the influence of
different parameters on the instability.
Rewritten in our dimensionless units as
\begin{equation} \label{condit}
\nu N > \frac{\rm v}{|{\rm v}'_E|},
\end{equation}
this inequality implies: (i) for fixed number of the SL periods $N$ there
should
exist the critical value for the doping $\nu^{*}$ above which instabilities of
the current may occur, and (ii) for any fixed doping $\nu$ one can bring the
system to instability by increasing the length of the SL $N$.
Precise bounds for  $\nu^{*}$ and  $\nu^{**}$ may be found
elsewhere.\cite{BKMW}


Notice the importance of the NDC region in the velocity curve $v(E)$ and the
positive sign of $\frac{\partial E}{\partial x}$ at the boundary, which are
both
necessary for monopole recycling.
Besides the oscillations of the domain wall due to the travelling charge wave,
the field in the low- and high-field domains also oscillates
[see Fig.\ \ref{En-it}(a)]. The field behind the monopole is uniform in space,
up to a small correction of the order of $\delta$ near the boundary. When the
current reaches its maximum then the field behind the monopole takes values
on the NDC region,
and those corrections increase exponentially in time, nucleating a new
monopole.
\cite{BKMV}

The condition (\ref{condit}) in the strong limit
$\nu N \gg {\rm v}/|{\rm v}'_E|$ implies a large separation between the
characteristic times $\tau_t$ and $\tau_{\epsilon}$. This can be exploited to
perform an asymptotic analysis of the current oscillations, which gives a
reasonable approximation to the numerical simulations for long SL's with
$N>100$. \cite{BKMV}
A key ingredient of the analysis is that the domain walls become shock waves
(i.e., discontinuities moving towards the right and separating quasineutral
regions of low and high electric field), whose velocity obeys an equal-area
rule.\cite{Bonilla-ICPF}

\section{Frequency-locking under dc and ac bias}

Self-oscillating systems that are forced by an external oscillating signal
represent an important class of coupled oscillators.
An inherent feature of periodically forced nonlinear system is that the actual
oscillation frequency depends on the amplitude of the forcing.
Therefore both the frequency  $f_d$ and the amplitude of the driving $a$ can be
used as control parameters to study nonlinear dynamics of our system.

For relatively low amplitudes we could expect two possible effects of the
forcing according to the ratio between the driving frequency $f_d$ and the
natural frequency $f_0$. If this ratio is a rational number the regime
of entrainment or mode-locking will be observed, if it is irrational the regime
of quasiperiodicity will occur because the frequencies are incommensurate.
\cite{Ott}

The statement above is confirmed by our numerical calculations.
The phase diagram of Fig.\ \ref{m-lock} shows the distribution of
frequency-locked solutions (Arnol'd tongues) over the frequency -- amplitude
parameter plane.
Within each tongue, the current is a periodic function of time whose actual
frequency $f_s$ does not coincide with $f_0$ nor with $f_d$ in general. Instead
it is usually related to one of their harmonics. Frequency-locking can be
considered as the trend of the driven system to keep its frequency close to
that
of the unforced oscillation, $f_0$, by taking the value of the harmonic of
$f_d$ which is closest to $f_0$. The periods of the actual oscillations are
shown by numbers in Fig.\ \ref{m-lock}. Notice that the frequency-locked
solutions are interspersed with regions of quasiperiodicity where both
frequencies $f_0$ and $f_d$ coexist.
At $a$=0 the Arnol'd tongues arise in intervals around rational ratios of
$f_d/f_0$, and here $f_s=f_0$. As $a>0$ increases, the tongues open up and
eventually intersect. Then more complicated dynamic behaviors (e.g., chaos)
usually occur as the solution jumps between the various overlapping modes in
an erratic way.

Frequency locking is a typical nonlinear phenomenon that can be
found in different periodically driven systems. \cite{Ott}
In our model the
Arnol'd tongues begin to overlap already at very low driving amplitudes
($a \sim 0.01$). This is quite different of the case of the Gunn diode studied
by Mosekilde {\it et al.} \cite{mosekilde} In addition, we observe 'the red
shift' in the dependence of the frequency $f_s$ with the amplitude $a$, which
manifests itself in the bending of the locked regions. The last phenomenon is
also absent in the bulk Gunn effect and it might be a consequence of the
delay of the natural oscillations when the forcing is increased.
We have a qualitative argument to understand why the overlapping of Arnol'd
tongues (and therefore the critical line for chaos) occurs at relatively low
ac-voltages.
Notice that our system has important voltage and time scales due to its
discreteness: recall that there are spikes superimposed to the oscillation of
the current tracking the jumps of the domain wall from one QW to the next one.
It is plausible that overlapping of Arnol'd tongues occurs when the ac voltage
amplitude is equal to the time-averaged voltage drop per period, $V_b/N$,
needed to balance the typical discrete scale of voltage.
The latter is equal to the energy separation between subbands divided by the
charge of the electron, ${\cal E}_{1-2} l = \Delta/e$.\cite{remark3}
Unlike the ac-driven bulk semiconductors, one could thus expect overlapping
of resonances for a SL driven with an ac amplitude $V_b A \sim \Delta/e \sim
V_b/N$, which is much smaller than the dc bias $V_b$. This value seems to
give a reasonable estimation of the lowest critical line for chaos as shown
in Fig.\ \ref{m-lock}.

Let us now further describe the Arnol'd tongue diagram of  Fig.\ \ref{m-lock}.
Between the main tongues correspondent to $n:1$ locking, smaller $n:m$
intervals
are found, representing more complex entrainments. For instance, between the
$3:1$ and $4:1$ tongues one can see $10:3$, $7:2$, and $11:3$ ratios, between
the $4:1$ and $5:1$ tongues there are $13:3$, $9:2$, and $14:3$ ratios, and so
on. Scanning over both parameters $a$ and $f_d$ to analyze the system of 40
equations was very time-consuming, because we calculated the largest Lyapunov
exponent (see below) to distinguish between quasiperiodic and chaotic
solutions.
Thus we restricted ourself to scanning with a frequency step $\Delta f_d = 0.05
f_0$, which allowed us to detect the tongues $n:m$ with at most $m$=3.
Higher order locking and much smaller new chaotic regions at the boundaries
of the tongues are expected to appear if a smaller step $\Delta f_d$ is used.
An additional important result in the mode-locking diagram of Fig.\
\ref{m-lock}
is that the richest structure of different mode mixing and chaos is
concentrated
between the $2:1$ and $3:1$ tongues. In the next section we consider in more
detail the results obtained when the driving frequency is fixed within that
region, namely, at the inverse golden mean ratio.

\section{Chaotic dynamics at the golden mean ratio}
\subsection{Bifurcation diagram}

Here we shall consider the driving amplitude $a$ as the control parameter,
fix the driving frequency as $f_d=G f_0$ (where
$G=(\sqrt{5}+1)/2 \approx 1.61803...$ is the inverse golden mean ratio
\cite{remark4}), and calculate the current $J$ as a function of time.

To detect and visualize the chaotic regions in parameter space, we need to
define a Poincar\'e mapping. The current is a good measure of the amplitude
(norm) of the solutions, which is illustrated by the use of current versus
voltage characteristics as bifurcation diagrams \cite{higuera92}.
Let $T_d=1/f_d$ be the driving period. We adopt as our Poincar\'e mapping (for
each value of $a$) the current at times $\tau_m=m T_d$, $m = 0, 1, \ldots$,
(after waiting enough time for the transients to have decayed). \cite{BB95}
The result is the bifurcation diagram in Fig.\ \ref{bifur}(a) which is
constructed as follows. For each $a$ we compute $J_m = J(m T_d)$ until the
solution becomes periodic within a $10^{-5}$ accuracy. At that time, we stop
the
simulation and depict all the $J_m$ corresponding to one period of the
solution.
If the solution is not periodic, we eliminate the first 500 transient points
$J_m$ and depict the next 200 points. Thus, aperiodic solutions can be very
easily distinguished from periodic ones in Fig.\ \ref{bifur}(a) by the large
number of points (periods) corresponding to each value of $a$. Notice that
period-2 orbits span the widest parameter region from $a \approx 0.01$ up to
$a \approx 0.085$.
Period-doubling cascades can be seen, which points out the existence of chaos
near their accumulation points.


The next important consideration is to discriminate chaotic from quasiperiodic
regions, both distinguished by having a large number of points in the
bifurcation diagram. This can be achieved by computing the largest Lyapunov
exponent $\lambda_1$. For chaotic regions $\lambda_1 > 0$, which indicates
exponential divergence of nearby trajectories. For periodic solutions of our
nonautonomous system $\lambda_1 < 0$, while for quasiperiodic solutions
$\lambda_1=0$.

In Fig.\ \ref{bifur}(b) we present the first and the second Lyapunov exponents
calculated by the algorithm explained in the Appendix.
We see that the system starts being quasiperiodic ($\lambda_1 =0$) at $a=0$,
as it should be for an irrational frequency ratio.
Then at $a \approx 0.005$ it locks to a period-5 orbit ($\lambda_1 <0$)
terminated by some chaotic windows at $a \approx 0.01$ ($\lambda_1 >0$).
Notice the first appearance of chaos at relatively small driving amplitude
($\sim 1\%$ of $V_b$).
After $a > 0.085$ several chaotic windows can be seen, and then the solution
becomes again quasiperiodic (as it was at $a=0$) before locking to the driving
frequency $f_d$ at $a \approx 0.145$.

More insight into the transition between chaotic and nonchaotic regions of the
bifurcation diagram can be obtained by sweeping-up (or sweeping-down) through
it, using the electric-field profile stored at the end of each simulation (for
$a=a_0$) as an initial condition for the next simulation (for $a=a_0+\Delta
a$).
This approach is close to the process of experimental sweeping-up measurements.
By this technique we investigated some critical points in the bifurcation
diagram and found different pictures when sweep-up and sweep-down runs were
made, thus demonstrating hysteresis.
For example, such hysteretic behavior we obtained for the transition point
between period-2 orbit and chaos at $a \approx 0.085$, which points out to the
existence of a subcritical bifurcation.

\subsection{Current-voltage phase plots and Fourier spectra}

Fig.\ \ref{phase} shows $I$-$V$ phase plots and the corresponding
Fourier spectra (FS) for some specific values of $a$.
These phase plots could be measured experimentally by depicting
the current in the external circuit as a function of the instantaneous value of
the ac $+$ dc voltage bias.
We observe different types of solutions as shown
in Fig.\ \ref{phase}. Periodic orbits appear as simple closed loops [Fig.\
\ref{phase}(c),(f)]; they have few frequency peaks in the FS, corresponding to
the driving frequency $f_d$ and its harmonics.
The quasiperiodic orbits look more complicated [Fig.\ \ref{phase}(a),(e)], but
their FS are still simple.
For the case of Fig.\ \ref{phase}(e) (strong driving) the main peaks appear at
$m f_d$ ($m=1,2,\dots$).
Additional small double peaks around $(m-\frac{1}{2}) f_d$ (with $m=1,2,\dots$;
the separation between the peaks of each pair is always the same for all double
peaks) are related to one of the harmonics of the natural frequency
$f_0$, thereby providing coexistence of the two frequencies $f_0$ and $f_d$.
For the quasiperiodic case of Fig.\ \ref{phase}(a) (weak driving) the main
peaks
are at $m f_0$ with the same type of the double peaks, which are already
related
to $f_d$.  Thus the natural and driving frequencies exchange their roles
depending on the strength of the forcing.
Finally, for the chaotic solutions [Fig.\ \ref{phase}(b),(d)] the FS become
very irregular with a large number of peaks, which could be considered as an
additional method to detect chaos.
The FS for chaotic solutions should not be necessarily continuous.
In our system the sharp frequency components are also present in the FS as in
the case of the familiar R\"ossler attractor. \cite{rossler}

The Poincar\'e mapping used to obtain the bifurcation diagram of Fig.\
\ref{bifur}(a) can be understood from Fig.\ \ref{phase} as the successive
crossing of the orbit through the line ${\cal V}$=1.2, where ac part of the
voltage crosses zero (at upper values of the current corresponding to the
increasing voltages).  For aperiodic solutions those crossing points are
distributed over some interval (or intervals) of the current.

\subsection{First return map}
\label{jmm}

One of the main quantities observable in experiment is the current density in
the external circuit $J(t)$.
By sampling the current $J(t)$ each driving period $T_d$, one obtains the data
set $J_m$. This set can be analyzed by means of the first return map
plotting $J_{m+1}$ as a function of $J_m$.
After some transient time the resultant attractor for a $n$-period solution
will
be just the $n$ separate points (0-dimensional object), while for an aperiodic
(chaotic or quasiperiodic) solution it will be represented by a higher
dimensional object.

A closed smooth loop with regular distribution of the points indicates
quasiperiodicity (Fig.\ \ref{Jm}a).
The chaotic attractor is a layered (sometimes folding) structure with varying
density of the points on different regions (Fig.\ \ref{Jm}b), and it can be
characterized by the multifractal dimension $D_q$. \cite{halsey}
Sometimes the chaotic attractor contains several separate branches almost
continuously filled as in Fig.\ \ref{Jm}c. In this particular case the
attractor
has five branches, indicating that the solution is close to period-5
locking for this value of the control parameter $a$. The chaotic solution has
the same symmetry as the frequency-locked solution which lies nearby.

\subsection{Spatiotemporal aspects of chaos}

So far we have only characterized the temporal aspects of our chaotic current
oscillations leaving aside the spatial dependence of electric field and
charge density. This dependence is a very characteristic feature of our system
because the oscillations of the current are due to the dynamics of
nonlinear travelling charge waves. Thus we analyze here what are the field
and charge density profiles in the dynamical regimes of interest described
previously.


Under pure dc voltage bias, the SL exhibits time-periodic current oscillations
accompanied by a periodical recycling of the monopole charge wave (domain wall)
in space as described in Sec.\ \ref{selfosc}.
When an ac signal is superimposed on the dc voltage bias, the current can
become
chaotic for particular values of the control parameters. For that case the
motion of the charge waves becomes chaotic too, as shown in Fig.\
\ref{ne-ch}(a). The ac part of the voltage causes the electric field in
different parts of the SL to take on values in the NDC region at different
instants of time. This results in irregular amplification or damping of the
charge disturbance at different QW's.


There are two significant new features for a SL under
ac and dc voltage bias as compared to the pure dc case.
The first distinctive feature is that the monopoles may be generated closer to
the beginning of the SL, thereby leaving more room to their motion towards
the end of the SL; see Fig.\ \ref{ne-ch}(a). Then the peak-to-valley ratio in
the current oscillations increases as it can be appreciated in Fig.\
\ref{phase}. Amplification of the ac signal is not linked to chaos since it
can also be observed in cases where there is frequency locking and an uniform
electric field profile inside the SL. \cite{amp} In fact, the amplification
is most pronounced at the instability threshold for the doping $\nu^*$, where
the current oscillations appear. Then the electric field is almost uniform all
the time.

The second new feature is the qualitative change of the traveling wave picture
for longer SL. For short SL ($N<80$), at most two domain walls (separating
three
domains) can coexist at a given time (see Fig.\ \ref{ne-ch}(a)), whereas up to
three coexisting domain walls (separating four domains) can be observed for
long
SL's during certain time intervals (for the case of Fig.\ \ref{ne-ch}(b) at
$\tau \approx 41.1; 46.5$).
In Fig.\ \ref{monop}(a)-(c) we show the presence of one, two or three electric
charge monopoles at different times for a 200-well SL under ac and dc voltage
bias. We can understand this as follows. The width of a domain wall ($\approx$6
wells for our particular set of parameters) is determined by the velocity
profile and it is approximately independent of the SL length. \cite{Bonilla94}
Then the monopole has more room to move on a longer SL (under pure dc bias,
monopole recycling takes place on about 12 periods of a 40-well SL as compared
to about 90 periods of a 200-well SL). The disturbance caused by the ac bias
on monopole motion is thus much larger for a longer SL, which results in quite
different spatiotemporal structures under ac and dc bias
[compare the charge densities of Figs.\ \ref{ne-ch}(a) and (b)].
For a short SL the chaotic behavior can be associated with a chaotic domain
wall dynamics that resembles the dc voltage bias case: most of the time there
is
only one domain wall which moves to the end of the SL and disappears, and
about that time another monopole is generated, sometimes quite close to the
beginning of the SL [see Fig.\ \ref{ne-ch}(a)]. On the other hand, for a long
SL
the graph of the electron density is disjoint:
In addition to long-living waves traveling over almost the whole SL, there are
short-living waves existing only at the beginning of the SL.
The two types of waves are distributed chaotically in space.


We have not found more than three monopoles by increasing further the
SL length $N$. This may be due to the fact that there is not enough charge in
the SL (determined by doping) to provide more than three jumps in the electric
field. It might be possible to observe more complicated field structures in
parameter regions where the current instabilities exist at much higher electron
densities.

The spatially chaotic nature of the solutions can be illustrated
by picking two far-away QW's and depicting the simultaneous values of
the electric field at them after each period of the driving force $T_d$.
The resultant attractors will be very similar to those obtained for the first
return map $J_{m+1}-J_m$ described in Sec.\ \ref{jmm}
(cf. Fig.2 in Ref.\ \cite{BB95} with Fig.\ \ref{Jm} of the present paper).

\section{Conclusions}

Dynamic properties of the high-field transport in weakly-coupled superlattices
under ac and dc biases has been studied numerically within the simple
self-consistent discrete drift model.
A nonlinear interaction between an internally generated periodical motion
of the accumulated charge wave and an external microwave signal gives rise to a
frequency-locking, quasiperiodicity or chaos depending on the external driving
parameters. The calculated phase diagram of the frequency-locked solutions
(Arnol'd tongues) shows the following new features:
(i) the first overlapping of tongues giving transition to chaos occurs under
very weak driving ($a \sim 0.01$);
(ii) there is a bending of tongues under strong driving ('red shift' in
frequency) which we associate with the delay in natural oscillation frequency,
and that is a consequence of our spatially extended system.

Besides the chaotic dynamics, the microwave forcing was found to give an
amplification of the self-sustained current oscillations, which is most
pronounced at the instability threshold for the doping $\nu^*$, where the
oscillations appear.

It would be of interest to verify our predictions by making measurements in
currently available $n$-doped GaAs/AlAs samples forming $n^{+}$-$n$-$n^{+}$
diodes. These samples exhibit self-sustained oscillations under pure dc voltage
bias, \cite{japan} and are thus suitable candidates for observing chaos when
an appropriate microwave signal is superimposed on the dc voltage bias. More
highly doped samples present multistable stationary solution branches
(corresponding to coexisting static electric field domains \cite{domains-exp})
in their current -- voltage characteristics under dc voltage bias. Studying
the response of these samples to ac and dc voltage bias would also be of
interest. Preliminary calculations show rich spatiotemporal behavior
including chaos.

\acknowledgements

We thank J.\ M.\ Vega, C.\ Martel, H.\ T.\ Grahn, and J.\ Kastrup for valuable
discussions. O.M.B. has been supported by the Ministerio de Educaci\'on y
Ciencia of Spain. This work has been supported by the DGICYT grants PB92-0248
and PB94-0375, and by the EC Human Capital and Mobility Programme contract
ERBCHRXCT930413.

\appendix

\section*{Lyapunov exponents}

Consider the $N$-dimensional phase space for the electric-field vector $E_i$.
The evolution of its trajectory $E_i(\tau)$ is described by the set of
nonlinear
equations (\ref{map}). Since we want to know how trajectories infinitesimally
close to the real trajectory of our system diverge from it, we can use a
linearized version of the equations (\ref{map}):
\begin{eqnarray} \label{lin}
\frac{d \hat{e}_i}{d \tau} = && \frac{1}{N}\sum_{j=1}^{N}
[{\rm v'}(E_{j}^{\tau}) \; n_{j}^{\tau} \; \hat{e}_j +
{\rm v}(E_{j}^{\tau}) (\hat{e}_j-\hat{e}_{j-1}) ] \nonumber \\ &&{} -
{\rm v'}(E_{i}^{\tau}) \; n_{i}^{\tau} \; \hat{e}_i -
{\rm v}(E_{i}^{\tau}) (\hat{e}_i-\hat{e}_{i-1}).
\end{eqnarray}
Here $\hat{e}_i$, with $i=1,\dots,N$, is the vector of the disturbances of
the electric field.
The velocity ${\rm v}(E_{i}^{\tau})$, its derivative ${\rm
v'}(E_{i}^{\tau})$, and the electron density $n_{i}^{\tau} = E_{i}^{\tau} -
E_{i-1}^{\tau} + \nu$ are calculated on the fiducial trajectory $E_i({\tau})$.
The boundary condition for the system (\ref{lin}) is $\hat{e}_0 = \hat{e}_1$.

To calculate the largest Lyapunov exponent $\lambda_1$ is very easy.
We take an arbitrary initial vector $\hat{e}_{i}(0)$. Then both the nonlinear
system (\ref{map}) and the linearized system (\ref{lin}) are advanced forward
in time simultaneously.

Denoting by $\|\hat{e}_{i}(\tau)\|$ the euclidean norm of the vector
$\hat{e}_{i}$ at time $\tau$, the largest Lyapunov exponent is given by

\begin{equation} \label{l1}
\lambda_1 = \lim_{\tau \to \infty} \frac{1}{\tau}
\ln \frac{\|\hat{e}_{i}(\tau)\|}{\|\hat{e}_{i}(0)\|}.
\end{equation}

The linear system without nonlinearities will give an exponential growth
(relaxation) of the solution for a positive (negative) $\lambda_1$.
Therefore, the perturbations will need to be renormalized from time to time to
prevent overflow (underflow), because they are only represented with a finite
floating point numbers in the computer.
To avoid this we use the algorithm of Benettin {\it et al.} \cite{benettin} and
renormalize our perturbation vector after each half period $T_d/2$ of the ac
voltage bias.  Then $\lambda_1$ will be given by

\begin{equation} \label{l1n}
\lambda_1 = \lim_{k \to \infty} \frac{2}{k T_d}
\sum_{j=1}^{k} \ln \|d_{j}\|,
\end{equation}
where $d_j$ is the vector perturbation growth during the $j$th renormalization
period.
Fig.\ \ref{lam1} shows the results for $\lambda_1(t_k)$ obtained after
computing
of $k$ renormalization intervals. In the limit $k \to \infty$ $\lambda_1$
converges to its limiting value (positive, zero, or negative) according to the
control parameter $a$.
Notice a little difference between $a$=0.01 and $a$=0.0102 gives a drastic
difference in values of the Lyapunov exponents from negative to positive.

To look at our chaotic solutions from the point of view of their dimensionality
we have to compute the next Lyapunov numbers.
Of course, we don't need all the Lyapunov spectrum of 40 numbers to
characterize
our system.
Kaplan and Yorke defined a
quantity called the Lyapunov dimension $D_L$, given by a formula \cite{kaplan}

\begin{equation} \label{dl}
D_L = K + \frac{1}{|\lambda_{K+1}|} \sum_{i=1}^{K} \lambda_i,
\end{equation}
where $K$ is the largest integer such that $\sum_{i=1}^{K} \lambda_i \geq 0$
(all $\lambda_i$ are arranged in decreasing sequence).
Since different Lyapunov exponents characterize the stretching and contracting
of phase space in different directions, the Lyapunov dimension is then the
number of vectors in phase space needed to describe an infinitesimal volume
that
remains constant on the average.
$D_L$ is related to the information dimension of the system $D_1$
\cite{kaplan}
and can be used to characterize the fractal dimension of the associated chaotic
attractor.


To calculate the first $n$ exponents $\lambda_1,\dots,\lambda_n$ we define $n$
initial perturbation vectors $\hat{e}_{i}^{(n)}(0)$, which are orthonormal, so
that they form $n$-dimensional sphere.  As the system is integrated forward in
time, the sphere of trajectories will become an ellipsoid since different
directions in phase space will expand or contract at different rates.
The Lyapunov exponents $\lambda_i$ are determined by the rate of expansion or
contraction of the principal axes of the ellipsoid averaged over the entire
attractor.
One additional problem arises here. Since all the vectors
$\hat{e}_{i}^{(n)}(\tau)$ tend to line up in the direction of the greatest
growth we have to use the Gram-Schmidt orthonormalization procedure to separate
the vectors into orthogonal components.
The Lyapunov exponents $\lambda_i$ are then calculated by the same formula
(\ref{l1n}) as for $\lambda_1$, with $d_{j}^{(n)}$ being the orthonormalized
perturbation growth of $\|\hat{e}_{i}^{(n)}\|$ during $j$th orthonormalization
period.

Fig.\ \ref{lam3} shows the temporal convergence of the first three Lyapunov
exponents for the particular chaotic solution ($a$=0.09).
The estimated from those values Lyapunov dimension $D_L \approx 2.208$
indicates low-dimensional chaos for chaotic dynamics in our SL.
By further increasing the length of the SL $N$ the characteristic Lyapunov
dimension does not grow. Since we did not see a transition from low- to
high-dimensional chaos, it could be concluded that our system is not
extensively
chaotic. We have not observed hyperchaos either, where the second
Lyapunov exponent is positive.

\begin{figure}
\caption{
Dimensionless effective velocity (proportional to the tunneling probability) as
a function of the electric field.  The point indicates the electric-field value
corresponding to the dc bias ${\cal V}=1.2$ used in the calculations.
}\label{v-E}\end{figure}

\begin{figure}
\caption{
Temporal evolution of the current $J$ (under dc bias) for different values of
the doping concentration $\nu$ indicated at the right margin with increasing
order from the bottom to the top. The boundary condition $\delta$=$10^{-3}$.
}\label{j-dop}\end{figure}

\begin{figure}
\caption{
Frequency of the natural oscillations $f_0$ under dc bias
(scaled on the total travelling time over the whole SL $t_{\rm tun} N$) vs
doping concentration $\nu$ for $N=40$ and for different boundary parameters
$\delta$. Out of the regions marked by dashed vertical lines oscillations
disappear.
}\label{f0-dop}\end{figure}

\begin{figure}
\caption{
Spatiotemporal distributions of electric field (a) and electron density (b) for
self-sustained oscillations under dc bias. Quantum well index is denoted by
$i$.
}\label{En-it}\end{figure}

\begin{figure}
\caption{
Phase diagram showing the distribution of frequency-locked solutions over the
driving frequency -- driving amplitude parameter plane.
Driving frequency $f_d$ is in units of the natural oscillation frequency
$f_0 \approx 0.0235 (1/t_{\rm tun})$.  Grey color corresponds to
quasiperiodicity, lighter color with different shades corresponds to periodical
solutions (number of periods is shown by numbers), chaos is marked by black
color.  $G$ is the inverse golden mean.
}\label{m-lock}\end{figure}

\begin{figure}
\caption{
Bifurcation diagram of the current obtained by means of Poincar\'e mapping each
driving-force period $T_d$ (a) and the first two Lyapunov exponents (b), both
versus the driving-force amplitude $a$ for the golden-mean ratio between
natural and driving frequencies. Windows of chaotic solutions are marked by
arrows.  Lyapunov exponents are scaled on the driving period $T_d$.
}\label{bifur}\end{figure}

\begin{figure}
\caption{
Phase current-voltage plots (left) and the corresponding Fourier spectra of the
current (right) for different driving amplitudes $a$: 0.005(a); 0.0102 (b);
0.05 (c); 0.09 (d); 0.144 (e); 0.16 (f).
}\label{phase}\end{figure}

\begin{figure}
\caption{
First return map for current density $J_m$ at different driving amplitudes $a$:
0.14(a); 0.09(b); 0.101(c), indicating quasiperiodicity (a) and chaos (b),(c).
}\label{Jm}\end{figure}

\begin{figure}
\caption{
Chaotic propagation of the charge accumulation waves in a 40-period SL for
$a$=0.09 (a) and in a 200-period SL for $a$=0.12 (b).
}\label{ne-ch}\end{figure}

\begin{figure}
\caption{
Nonuniform distribution of electric-field domains for a 200-period SL:
(a) $a$=0 at the time moment $\tau=3.20$ (one monopole);
(b) $a$=0 at $\tau=4.35$ (two monopoles);
(c) $a$=0.112 at $\tau=18.05$ (three monopoles);
}\label{monop}\end{figure}

\begin{figure}
\caption{
Temporal convergence of the largest Lyapunov exponent $\lambda_1$ for different
driving amplitudes $a$: 0.01 (periodic); 0.0102 (chaotic); 0.05 (periodic);
0.1395 (quasiperiodic).
Time $t_k=k T_d/2$ is in units of the renormalization intervals.
Lyapunov exponents are scaled on the driving period $T_d$.
}\label{lam1}\end{figure}

\begin{figure}
\caption{
Temporal convergence of the first three Lyapunov exponents for chaotic
solution at $a$=0.09.
Lyapunov exponents are scaled on the driving period $T_d$.
}\label{lam3}\end{figure}


\begin{thebibliography}{99}
\bibitem{remark2}
These resonances are purely quantum mechanical phenomena that have to be
included in our otherwise semiclassical model.
In our calculations we use the same curve $v(\cal E)$ as in Ref.\ 14
which corresponds to the experimental current-voltage characteristic of
Ref.\ 8.
\bibitem {remark3}
This quantity measures the voltage drop per period when the whole SL is
undergoing resonant tunneling. It is also the voltage spacing $\Delta V$
between
solution branches in the stationary $I$-$V$ characteristics of a SL in the
regime of stable domains \cite{Prengel,Bonilla94,NATO}.
(Actually $\Delta V$ is approximately $\Delta/e$ minus the half-width of the
corresponding resonance peak, therefore slightly smaller than $\Delta/e$).
\cite{zhang} ${\cal E}_{1-2} l$ is also the voltage separation between the
regular peaks in the capacitance-voltage characteristics. \cite{zhang,liu}
\bibitem {remark4}
The golden mean is often used in driven nonlinear systems because of its number
theoretic properties. Specifically, in some sense, it is the most irrational of
all irrational numbers in that it is the most difficult to approximate by
rational numbers of limited denominator size. \cite{Ott}
\end{thebibliography}
\end{document}